# Interface between market and science

At the beginning, programming was inspired by the search of the best solutions.  At that time some fundamental stones like famous languages and object oriented and structured programming were laid.  It was found later that applications could generate huge profits; after it marketing departments started to decide what was right and wrong.  Programs are ruled by developers but declared user-friendly; millions of users are going mad trying to get the needed results from these applications.  Research goes on and new results can be opposite to business view.  History shows that not science has to adjust to business, but eventually business will have to adapt to the results of the research work.

**Introduction**

In our days nearly half of the Earth population presses the keys of some personal computer or a smartphone.  Each press is transformed into some action by one of the programs.  Pressing a keyboard or clicking a mouse does not differ from pressing some button on TV remote control or on the car panel, but there is a fundamental difference in results.  With the cars, TV and radio sets, washing machines and other home devices, we get the expected reaction.  With the personal computers you never know; it can be an expected one, or close to it, or far from expected, or there can be no reaction at all.  Such variety of results is caused by one thing: in computers, not the direct action but usually an interpretation is used and this interpretation was programmed by someone with different ideas about the expected results.  The use of interpretation is based on the history of programming, on the differences in priorities of the involved parties (researchers, programming companies, and users), and on one wrong step that was made many years ago and which eventually brought us to the current situation.  Without a brief look into the history of interface design, it is impossible to understand the core of the problem and the new ideas.

**There was a period without tensions between users and developers**

60 years ago computers were rare but big; each one occupied not a small room and required specialists for maintenance.  Soon the sizes of computers were reduced but their main purpose was still the same: they were very powerful calculators for solving engineering and scientific problems.  Specialists working on engineering problems were proficient in math and had no problems with learning programming languages.  A lot of scientists were writing programs for their own work.  The program results were printed out.  When programmer and user is the same person, then there are no conflicts about output.  Results of any program were shown in numbers (scalars and arrays) and there were no interface problems.  There were restrictions on computer memory and solving of some problems required hours of calculations, so there were huge efforts to use less memory and to invent the fastest algorithms.  Required memory and calculation time can be easily checked, so the comparison of those algorithms was very objective.  Money did not play any role in the best algorithm selection and a lot of exceptional results were obtained by university researchers.  It was the Golden Age of programming as a science; at that time famous languages were designed and fundamental rules of programming were formulated; a huge number of publications were devoted to the scientific aspects.

Personal computers were invented for profit.  They turned out to be very profitable but for this they had to turn the whole computer world upside down.  In our days PCs are still used to solve engineering and scientific problems but due only to the progress of element base; modern PCs can be used for calculations which were problematic even for big computers 30 years ago.  The majority of engineering and scientific problems are now solved on PCs, so there are still programs of such type, but their part among all applications is so small that they play the role of Cinderella long before happy end.  Throughout their history, PCs are targeting not the problems of some type but rather groups of people.  There is also the constant attempt to widen those groups because the profit correlates with the number of users.

**History of interface started with PCs**

At first PCs worked under DOS, but even at that period the emphasis shifted from calculations to interface.  The number of users greatly increased though they still represented the scientific and engineering world.  Those users could transform their algorithms into code, but the high-rocketing interface possibilities (especially after switching to graphical mode) and specific problems of this area caused the appearance of specialists in programming and the rapid growth of their number.  Not being specialists in any particular area of science or engineering, programmers know how to deal with computers, how to translate users' requests into machine codes, and how to show the results of calculations in different ways.  Initially the appearance of programmers was an excellent solution because it took out from many users the burden of learning those things which they didn't want to spend their time on (interface, computer graphics, memory usage), but eventually the mandatory presence of interpreter between users and computers grew up into a huge problem.

The spread of multi-windows operating systems changed the way of dealing with computers and set the rules under which we still work.  Millions of users can work with computers only if all the programs use the same (and few) easy to understand and deal with screen elements.  Such elements were introduced and their set hardly changed at all throughout the last 30 years.  There are buttons to be clicked with a mouse; there are lists with the strings to be selected by a click; there are special edit boxes into which a text can be typed from the keyboard; there are areas to show pictures.

For developers, all these elements are known as controls; for users they are simply the screen elements with easily remembered and used commands. These commands include scroll of the shown information and selection of the needed part; they also include the use of single and double click with possible difference in reaction for left and right mouse buttons. Some commands are associated with one or another screen element; the need of more general commands resulted in invention of menus. Nearly simultaneous introduction of menus and controls with all their possibilities caused an outburst of ideas in interface design. There were thousands of articles and hundreds of books. The work on new languages and better algorithms never stopped, but publications on interface absolutely prevailed.

Complex programs have to provide users with a lot of possibilities, so the number of available commands grew immensely. Developers tried to save users from sinking in that ocean of possibilities by proposing what they called "natural commands". In reality there is nothing natural in using a mouse with two buttons and setting the difference for single and double click. All those commands are not natural but easy-to-use-and-easy-to-remember; only such commands won the competition and are used. There is only one really natural action. Surprisingly, it is used only on a very limited basis, but there is a very serious reason for this. Better to say that there WAS a serious reason with a lasting effect.

**Dialogue at two levels**

The multi-windows operating system organizes dialogue of users and computers on two levels. At the upper level which can be also called the level of operating system, all programs are available and each one can be shown in two different ways. When the program is not started, it is shown as a small icon. Its picture is placed on a small rectangular area with invisible border; there is the same reaction on the click inside the whole area, so in reality we have a rectangular icon. This small nonresizable rectangle can be moved around the screen with a simple chain of actions: press at any inner point – move – release. There is a lot about **movability** in this text; under this term I mean the movement of an object not according to some predefined scenario but only by the direct user's action.

Working program is shown as a bigger rectangle with a border and special strip on the top (title bar); the remaining inner area is occupied by some needed elements. Big rectangles can overlap and their order of appearance (of overlapping) is decided by the simple clicks. Area of working application is resizable by border and is movable by the title bar. Thus, at this upper level all objects are movable. Fixed sizes of small icons do not matter at all because they are really small, while the sizes of the big rectangles can be easily changed. By using only two commands – move and resize – we can rearrange the view at the upper level to whatever we want. <u>User is the absolute ruler of the screen view at the upper level</u>.

There can be a significant number of icons on the screen; it is rarely less than 20, usually there are between 40 and 50 icons, and often enough the number can aspire to 100. Yet, no one discourse about the users' inability to deal with such number of icons or with the placement and size of rectangular areas. Users are allowed to organize them in the way they prefer. Developers of the operating system are responsible for easy moving and resizing of those objects and guarantee that the double click on any icon starts the associated application. There is a perfect division of roles: developers provide the correct work while users have the full responsibility and total control over the view. Developers are ready to help with this process but only to help and only after user's request.

Upper level is important but it is an auxiliary level. The real work is done inside the applications. By starting any program, we go to the second (inner) level and have to deal with the elements inside some rectangular area. There are some similarities with the upper level, but there is also one important difference.

When I was introduced to the Windows system by one of my colleagues, he showed me the icons movement, then started one application, and demonstrated the moving and resizing of its rectangular area. I was sure that everything had to work in the new exciting way, so I asked about the movement of the inner elements and was slightly puzzled by the answer: "No, it is impossible". It looked a bit strange to me because it marred a nice picture of the new exciting world, but at that time I worked under DOS where everything was fixed, so this immovability of the inner elements did not look like a serious flaw to me. It would be good to have everything movable, but if it is forbidden, I can live without it. Only years later I began to think more about the strangeness of that situation.

The inner area of any program can be populated by elements of two different types: controls and graphical objects. Controls are similar in behaviour to rectangles of the upper level. All standard controls are rectangular; they do not have title bar but they have borders. Each control has a set of methods to fill it with information and to get this information. Each control is provided with a set of methods to react to the clicks inside their area; those reactions are well known to developers and users, so no developer would think about substituting those reactions with something entirely different. In this way all controls have standard behavior in all programs and this makes the use of all applications much easier. Graphical objects are born by the skill and imagination of developers all round the world, so their shapes and sizes are unrestricted. Now we come to the problem for which good solution was neither introduced 30 years ago nor later and with which we all – developers and users – have to deal up till now.

**One old decision and the sequence of forced steps**

All controls are designed movable, but this feature is never used in applications because… graphical objects are unmovable. It is impossible to populate some area with a mix of movable and unmovable objects because this will produce a real mess; it would be like placing big stones on the highway. Either all elements inside an area are movable or none; developers perfectly understand this situation and never use possible movability of controls in applications. Users are never allowed to press control and move it.

Making any object movable is a serious task. Developers of multi-windows systems solved this problem for the easiest case – for rectangles – so all elements of the upper level and controls on the inner level are designed movable. There were several programs in which the movable objects of nonrectangular shape were introduced. The best known of such programs is *Paint* and there are others in which the movability of some elements is the mandatory thing. There are only few programs of such type and each of them provides the movability only for elements which are used inside this particular application.

Providing movability for specific objects inside application is a very skillful work and it was done in several programs. An easy to use algorithm of turning an arbitrary object into movable is a problem of much higher level. Multi-windows operating systems were developed by exceptionally talented people. I am sure that they thought about such algorithm because it was an obvious thing to be done. If such algorithm was not introduced, then it means that developers of those systems didn't know how to do it. The lack of such algorithm 30 years ago has long lasting effect (up till now) and, as often happens, one wrong decision causes not good decisions further on. In chess it is called a forced move.

The immovability of graphical objects causes the use of fixed controls, but sophisticated applications have to provide variants with different sets of inner elements and variations of their positioning. If users are not allowed to move and resize anything, then developers have to do it. Developers also have to provide interface through which users can send their requests for such changes.

Programs became more and more sophisticated and gave users a lot of possibilities; this could go only in parallel with introducing users to wider set of commands. The *adaptive interface* was announced as a perfect instrument with which users could adapt the designed program to their personal demands. In reality it means (but is never declared!) that developer provides several variants which are good enough from his point of view; users are allowed to select only between these variants. It doesn't mean at all that user needs exactly one of these variants, but he is not asked about his own desire. This problem was perfectly described in preface to [1]. *"So-called "applications" software for end-users comes with an impressive array of capabilities and features. But it is up to you, the user, to figure out how to use each operation of the software to meet your actual needs. You have to figure out how to cast what you want to do into the capabilities that the software provides. You have to translate what you want to do into a sequence of steps that the software already knows how to perform, if indeed that is at all possible. Then, you have to perform these steps, one by one."*

To organize placing and resizing of the screen elements, programmers were introduced to anchoring and docking. These instruments allow to fix the spaces between inner elements and window sides or to change sizes of elements linearly with the window change. It can be acceptable for the case of one, two, or three inner elements; it is bizarre for bigger numbers (and in normal applications you have 20, 30, or more elements inside), but it is an absurd from the point of common sense: user changes the application size in order to have a better view, but somebody else decides about the change of the elements inside. It is that interpretation which I mentioned in introduction.

Suppose that there are only two elements inside window. You are not allowed to change them directly, so you resize the window in order to… By the way, what was your idea about the needed changes inside? Do you want to change one object without changing another? Which one has to be changed? Do you want both to change in the same way or do you want to change their sizes in different ways? Where and how do you want to declare all these things? These are variants for the simplest case of two elements. If you are a programmer, are you ready to organize and introduce to users a system for positioning and resizing 30 elements? If you are a user, are you ready to learn and remember such system of commands?

In no way the standard interface would allow to solve this complex problem, so somewhere 10 years ago the *dynamic layout* was introduced and declared as design philosophy for the future. It strips users of any control over inner view of an application because designer decides about it for any window size. Surprisingly, this was announced as a greatest step in interface design because everything was done to simplify users' life. There are no commands to be remembered; user has nothing to worry about; simply resize the window, get the result, and applaud the designer. It reminds me of one similar situation: slaves do not need to worry about food because they are provided with it. Does it mean that slaves have to admire their life?

Each new "great achievement" in interface of PCs made this interface more and more user-friendly by leaving users with less involvement in making any decisions. Looks like programming companies decide that users are too stupid to make any decisions of their own. What is the users' role in such model? To pay regularly for the new version of some sophisticated toy and to applaud its creator?

Before looking at further results in the area of PCs, let us have a glimpse of the closely related but slightly different area. There is a wide variety of devices with relatively small screen. The main instrument of user's control over such device is finger. This instrument provides a limited set of commands: click the needed element to start the associated program, scroll the whole area, or zoom it. It turned out that for the most popular programs which are used by millions of people this set of commands is just enough. The question was about profit; the interface simplification immensely increased the number of users. Interface of all those devices uses the direct action. There are no words like "you are too dull (or too inexperienced) to do such thing, so we'll do it for you". The results of one's work depend only on the user's skills, so more skillful user gets better results in a shorter time. Now let us return to the world of PCs, but keep in mind the good features of interface on tablets and smartphones.

**Feeling of stagnation and the search for its root**

Throughout my whole career I was involved in development of very sophisticated programs for different branches of science and engineering. I did it for my own research work and I did it for others. Areas were far away from each other (applied optimization, voice recognition, telecommunication, big electricity networks, thermodynamics, and some others), but many years ago I came to the conclusion that, regardless of the area, the development and use of those programs went along the same lines while the differences were minimal and unimportant. I like to work with talented people and it happened so that I often had to discuss problems with such people. All of them worked on very serious problems and their research work required applications of not ordinary type. I mention this in order to explain the situation which pushed me into thinking not about the small changes for the new version of some program but about the solutions which would allow to design programs of another level.

What is the normal life cycle of a big and widely used engineering program? Once or twice a year users discuss with developers problems of the current version and make their suggestions about the needed changes. Users can be the best specialists in their specific area and usually work on the most difficult problems which, before they are solved, can be discussed only with some level of uncertainty. Developers have some level of understanding in this area, but they are lesser specialists. (Developers do not like this statement at all, but no one except them argues with this fact.) Developers listen to professional explanation of the problems but understand them only to their own level of knowing this area. Then they return to their office and start working on improvements according to their level of understanding. They are doing their best as developers (in this they can be the world class specialists), but in the specific engineering or scientific area they are not the best and cannot jump above the head.

"Generals always fight the last war" and these generals (in our situation – developers) are even not the best. Those programs are steadily changed (certainly improved!) from version to version, but this is going inside the best (the most profitable) situation for developers. Users pay for each new version; developers have the constant income.

As a result, there exists only one and up till now not questioned by anyone model of programs development. Any sophisticated (and not so sophisticated) application is developed by some group of specialists in programming who are, to some extent, familiar with requirements and needs of one or another special area. From time to time users complain about some problems; their requests are estimated and either fulfilled or rejected if, from developers' point of view, those request are too revolutionary. Because the whole process is organized for decades in such a way that any changes in interface can be authorized and implemented only by developers, then their opinion is final without any place to appeal. Nobody is thinking about appeal because it is going in this way for so long (forever on the time scale of computer world) that it became an axiom long ago.

This whole situation reminds me one well known story. For 15 centuries the Earth was considered by everyone to be the center of Universe. It was an axiom not to be discussed. Then first one and then another started to write about different view. Eventually and as the result of studies, calculations, and experiments that axiom turned out to be absolutely wrong.

Somewhere 12 – 14 years ago I got a strong feeling of stagnation in design of complex engineering programs. That feeling was not born spontaneously and it was not limited to some specific area. While I worked on a big project in a company, I was too busy with the current work, but the problem of that stagnation had to be explored. In 2006 I put myself into the place and situation where no one would disturb my thoughts and started to work on this problem.

**New features in one familiar program**

For better understanding of programming problem, let us consider one well known task. Suppose that you want to look at a function described by Y(x) equation. You need an application in which you can type this expression and declare the interval for argument, then the built in interpreter calculates this expression for an array of arguments and the program draws a graph. I think that Excel can do it as well as other programs. There is absolutely nothing new in writing such interpreter; it is an exercise for students of programming courses who are introduced to stacks and queues; I saw the explanation of such interpreter in the book years ago. Drawing a Y(x) graph based on two arrays is also an easy exercise for CS students, so instead of Excel you can write your own program.

Now suppose that you want to see and compare several graphs shown in separate areas. The same procedures for calculation and drawing, but there is an immediate growth of interface complexity. The number of graph areas must be declared and their sizes and positions must be chosen. The number of areas can be typed inside edit box, but there are variants for sizes and positions. Do those areas have to have equal sizes? Do they have to be lined along rows and columns or is an arbitrary positioning allowed?

Similar task of positioning and resizing for a set of elements can be seen in special systems for program design; Visual Studio is a good example. In this case the interface for positioning and resizing is provided by excellent specialists (maybe the best), but it is so intricate that only devoted programmers can use it. Implementation of the same interface in a program for graph analysis would be ridiculous and nobody is going to use such application. However, if users can move and resize those areas in the graph analyser easily and quickly, then you simply eliminate all those interface problems for declaring sizes and positions. You get not the simpler version of some complex interface; you don't need anything additional at all; only the graphical areas remain. Thus, only one SMALL problem is left: you need movable graphical areas.

A skillful developer can provide movability to the objects of the known structure. If you design a function analyser and you are good enough to provide movability to graphical areas, then you can do it and decide that the case is closed. That's what happened with the *Paint* program and several others, but I had different view on the problem.

**Find the problem, roll back, and start in the new direction**

Movability of objects is the only really natural command in interface. In our every day life, whenever we think that some object will suit us better if placed differently, we move this object without any second thought. Had you ever a feeling that you would prefer to place the elements inside application in a different way? I often want to do it. There are applications which provide variants (adaptive interface became a religion), but they never give me what I want.

*"In science, finding the right formulation of a problem is often the key to solving it..."* [2]. When I started to work on the problem of stagnation in scientific and engineering applications, I quickly understood that it was caused by the developers' ruling over all programs. While thinking about the ways to switch balance to users' side, I asked myself a simple question: "What would happen if graphical areas would be movable and resizable?" There was only one way to answer this question – to invent an algorithm for doing it. I started experiments on graphical areas, but this was due only to the fact that my main interest is in design of engineering and scientific programs. All such programs use graphs, so they were the most obvious elements to begin with. I was not working on an algorithm to be applied only to graphical areas of special class but on an algorithm that:

- Could be applied to objects of an arbitrary shape
- Would be easy to use by any programmer

Algorithm allows to move and resize arbitrary objects but doesn't dictate the way to do it. Moving by any inner point and resizing by border looks like the most obvious way to me, so this is implemented in all my applications. Users are not going to read instructions for each element, but knowledge of these simple rules allow them to deal with any program.

First I applied movability to the pure graphical areas; then there were several forced steps (they are described in details in my book [3]), and shortly I came to the idea of ***user-driven applications***. What makes them different from currently used programs?

My goal in design such algorithm was very simple and at the same time very ambitious: if any graphical object can be easily turned into movable / resizable, then I can return years back, change the underlying stone of interface design, and start the construction of entirely new building. Well, at first I had no idea that the building would be absolutely new. I had 25 years of experience in development of sophisticated programs and it was not so easy to get rid of my habits in design. At the same time I have a solid academic background (prior to my work as a programmer in business, for many years I was a researcher in the Academy of Sciences) and I know one thing very well: if the results of experiments show the discrepancy with theory, then there is some problem with this theory even if it is popular and acclaimed.

I have used the ideas of adaptive interface for many years nearly automatically and when my testing programs began to demonstrate that movable objects could not coexist with the ideas of adaptive interface, then at first it was not easy for me to put aside those well known rules. It was really an interesting experience when I was implementing movability in several of my older good working applications and regardless of the complexity and the purpose of those applications I got the same result. Each time there was such a sequence of steps.

- I took an application and turned into movable its main (crucial) object.
- It was a huge improvement in the work of this particular object but this caused a mess with unmovable objects around it.

- I turned those contacting unmovable objects into movable. That increased the set of perfectly working objects and moved the border of the mess slightly aside because there were still the unmovable objects farther on.
- In order to prepare a good working program, I had to repeat the previous step until ALL objects became movable.

This standard pattern resulted in the formulation of the main rule of user-driven applications: **All elements are movable**. All means really ALL! Programs vary from simple to extremely complex. The Demo program of my book contains 185 different windows (forms) and all elements in all these windows are movable. It is really the World of Movable Objects.

There are several more rules of user-driven applications, but they are closely related to the first one. There is one very interesting consequence of using these rules: with all elements movable, there is no more sense in division of elements on controls and graphical objects.

The compressed history of interface and the alternative way can be formulated now in several sentences.

- <u>Lack of an algorithm to move graphical objects</u> causes the use of controls only as unmovable; users are not allowed to change the view directly. Users ask for needed view changes and developers do something in an answer (*adaptive interface*). Such interface became so complex and inadequate that users were simply excluded from this process and now everything is decided by developers (*dynamic layout*).
- <u>Easy to use algorithm makes all graphical objects movable.</u> With all elements movable, users can change the view in any way and there is no need to discuss it with developers. The entire control over programs is passed to users (*user-driven applications*).

Programmer is responsible for good default view and provides the correct work regardless of users' actions. 30 years ago similar things were organized on the upper level of multi-windows operating systems but never spread farther on; with the invention of the algorithm for objects movability, I found the way to organize such interface in any program.

**From experiments to practice**

If CS is a science then fundamental laws of any science must be applicable. This means three things.

- Theory must be based on facts and experiments. As was formulated [4] long ago by British scientist William Gilbert: "*In the discovery of secret things, and in the investigation of hidden causes, stronger reasons are obtained from sure experiments and demonstrated arguments than from probable conjectures and the opinions of philosophical speculators.*"
- Experiments must be described in details so that it would be easy for anyone to repeat them and compare his own results with the announced.
- Anyone can organize his own experiments in order to approve or refute the proposed theory.

In CS the most widely used experiments are programs, so my algorithm for objects movability and the idea of user-driven applications are backed with a lot of programs. There are nearly 200 examples in my book; codes for all of them are available. These examples vary from the simplest, which are used to explain one or another detail of algorithm, to very complex which demonstrate the full power of user-driven applications. Those programs are also from very different areas. Here I am going to mention several examples, some results of their use, and a couple of conclusions from this experience.

Throughout the period of 2008 – 2012 I worked in the Department of Mathematical Modelling of the Heat and Mass Transfer Institute. Research in the area of thermodynamics is based on a lot of experiments. The results of experiments are used as the data for serious calculations, but this data is obtained together with the noise. Standard methods of calculations go crazy when they are used on the data with noise, so a preliminary data refinement is needed. This is a manual process which looks at the least ridiculous if organized with a standard interface; in user-driven application it is designed without problems and looks very natural. New type of applications perfectly suits to situations which can be hardly described by the strict rules and where some manual and non-standard user actions are required. The more manual actions are needed, the more are the advantages of user-driven applications.

Physicists especially asked me to work on the programs about which they thought for many years but which could not be designed in an old style. Under traditional interface, such programs would be so cumbersome that it would make them impossible to use. As user-driven applications, they are elegant, easy to use, and very useful for research work.

Several examples with manual data refinement are demonstrated with the book, but I want to underline that this is not some specifics of thermodynamics. While writing this article, I had a discussion with a researcher from telecommunication company. They have to do the same refinement of their data, but because they have no programs of the new type, they are forced to go through a very tiresome process with Excel.

To show possible changes in the well known program, I demonstrate a familiar *Calculator* but with movable elements. I do not see why everyone has to use the predefined view of this program. Mouse click on each screen button produces a definite reaction and this reaction does not depend on the size or place of a button. If you move elements around the screen and place them in any way you want, the Calculator still works as usual, but you have the preferable screen view. Try not to react with the words like "nobody needs it because Calculator is perfectly designed and suits everyone in the way it is". Lets everyone speak for himself. Just decide for yourself whether you prefer the fixed design enforced on you or would you like, in addition to the same perfect design, an easy and quick (just seconds) way to change the view to whatever you want at any moment. Maybe you would like to place ten buttons with numbers in one row, or in two rows, or around the circle as in old phones. It is a program on your computer; you can do with it whatever you want; at any moment you can return to the default view with a single menu command.

I am a bit surprised by the level of interest to my *Family Tree* application. There are several programs of such type and I wasn't thinking about developing a concurrent program. I was thinking about different types of block diagrams with rectangular elements connected by broken lines. Depending on some additional restrictions, such set of elements can be used in different areas and for different purposes. At one moment I understood that such set could represent a family tree. In my Family Tree, there is only one restriction from the point of Biology – you cannot be declared a parent of yourself. Everything else is allowed; you are free to organize any set of relations and position the elements in any way you want as all of them are movable.

Programs from different areas can be very much alike in elements and behaviour. Family Tree view is close to block diagrams and there are two different examples with such diagrams. I had a brief discussion with one reader about an application to be used in the university course on linear circuits. My advice was: take the codes of Family Tree example and organize similar connections not between rectangles but between elements of three or four different types.

How easy is the use of the proposed algorithm? One reader used simple examples from the beginning of the book to develop an application which helped him to position the furniture in his restaurant. If I am not mistaken, the place is somewhere in northern Italy. Maybe one day I'll have a chance to estimate the practical results of my ideas.

**Conclusion**

Adaptive interface is the dominating (or the only) paradigm of interface design for currently used programs. Though it is declared as a user-friendly interface, it is absolutely ruled by developers, so it suits programming companies very well. Nobody asks users whether it suits them or not. Users are asked only about the problems of particular application and if the problems are too big, then the new version is provided. The question of switching control over programs from developers to users is never discussed.

However, the popularity of tablets, smartphones, and other gadgets demonstrates the possibility of programs with direct users' control over the screen elements. Programs on those devices are relatively simple and the set of commands is very limited, but the devices are popular and there are no questions about users' ability to work with such interface.

Programs on PCs are more sophisticated and often use much wider set of commands. Such applications can be ruled by users only when all the screen objects become easily movable and resizable. I have designed such algorithm, but this is only an instrument of total movability. Algorithm to be used can be this one or any other; regardless of the method with which all objects are turned into movable, the programs of the new type (user-driven applications) will work under the same rules because all these rules are simply dictated by movability. The advantages of the new design increase with the growth of application complexity. The new ideas are extremely valuable for the cases when the needed changes cannot be strictly formulated and elements require the manual adjustment.

I demonstrate that any screen object can be easily turned into movable. There is no problem with applying the demonstrated (or similar) algorithm. The problem can be with programming companies which can be afraid to pass control over applications to users. Developers' apprehension can be lessened by the fact that tablets, smartphones, and other gadgets already moved in this direction and demonstrate good results. I am sure that PC users will get the full control over applications and this will change the process of our work on computers. The question is only about the moment when this will happen.